\documentclass[aps,amsmath,superscriptaddress,longbibliography,prx,twocolumn,footinbib]{revtex4-2}
\usepackage{graphicx}
\usepackage{dcolumn} 
\usepackage{relsize}
\usepackage{anyfontsize}
\usepackage{xfrac}
\usepackage{bm}
\usepackage[utf8]{inputenc}
\usepackage{amsmath}
\usepackage{amsthm}
\usepackage{hyperref}
\hypersetup{colorlinks,allcolors=blue}
\usepackage{physics}
\usepackage{mymacros}
\usepackage{natbib}
\usepackage{amsfonts}
\usepackage{tikz,pgf, tikz-3dplot}
\usepackage{tabularx,stackengine}
\usepackage{soul}
\newtheorem*{conjecture}{Conjecture}
\usetikzlibrary{decorations.markings,decorations.shapes,decorations.pathmorphing,svg.path, 3d}
\usetikzlibrary{chains}
\renewcommand{\selectlanguage}[1]{}

\newcommand{\rpsi}{\psi^{\otimes 2r}}
\newcommand{\ncp}[1]{^{\otimes #1}}
\newcommand{\Z}{\mathbb{Z}}
\newcommand{\per}[1]{\pi_{(#1)}}
\newcommand{\fabc}{\varphi_{\a\b\g}}
\definecolor{darkred}{HTML}{D40000}
\definecolor{lightblue}{HTML}{0066FF}
\definecolor{darkblue}{HTML}{003399}

\begin{document}
\tikzset{arr/.append style={
        decoration={markings,
            mark= at position {#1} with {\arrow{>}} ,
        },
        postaction={decorate}
    }
}

\tikzset{midarr/.append style={
        decoration={markings,
            mark= at position .5 with {\arrow{>}} ,
        },
        postaction={decorate}
    }
}

\tikzset{%
  dots/.style args={#1per #2}{%
    line cap=round,
    dash pattern=on 0 off #2/#1
  }
}

\newcommand{\origami}[2]{
    \begin{scope}[shift={(0,0.866025404*#1)}]
       \foreach \x in {0,60,...,360}{
           \draw (0,0) -- (\x:#1);
           \draw (\x:#1) -- (\x+60:#1);
       }
       \draw[decorate,decoration={zigzag,segment length=3pt,amplitude=1pt},thin] (-56:#2) -- (-5:#2);
       \draw[decorate,decoration={coil,segment length=3pt,amplitude=1pt},thin] (60-5:#2) -- (5:#2);
       \draw[dotted] (65:#2) -- (115:#2);
       \draw[dashed,thin] (125:#2) -- (175:#2);
       \draw[dash dot] (182:#2) -- (238:#2);
    \end{scope}
    \begin{scope}[shift={(0,-0.866025404*#1)}]
       \foreach \x in {0,60,...,360}{
           \draw (0,0) -- (\x:#1);
           \draw (\x:#1) -- (\x+60:#1);
       }
       \draw[decorate,decoration={zigzag,segment length=3pt,amplitude=1pt},thin] (56:#2) -- (2:#2);
       \draw[decorate,decoration={coil,segment length=3pt,amplitude=1pt},thin] (-58:#2) -- (-2:#2);
       \draw[dotted] (-62:#2) -- (-118:#2);
       \draw[dashed,thin] (-122:#2) -- (-178:#2);
       \draw[dash dot] (-182:#2) -- (-238:#2);
    \end{scope}
}

\newcommand{\pizza}[3]{
\begin{tikzpicture}[every node/.style={font={\scriptsize}}]
    \def\r{.4};
    \draw (0,0) circle (\r);
    \draw (0,0) -- (-90:\r);
    \draw (0,0) -- (-210:\r);
    \draw (0,0) -- (-330:\r);
    \node at (90:\r/2) {$#1$};
    \node at (-180+45:\r/2) {$#2$};
    \node at (-45:\r/2) {$#3$};
\end{tikzpicture}
}

\newcommand{\tpizza}[3]{
\vcenter{\hbox{\begin{tikzpicture}[every node/.style={font={\tiny}}]
    \def\r{.3};
    \draw (0,0) circle (\r);
    \draw (0,0) -- (-90:\r);
    \draw (0,0) -- (-210:\r);
    \draw (0,0) -- (-330:\r);
    \node at (90:\r/2) {$#1$};
    \node at (-180+45:\r/2) {$#2$};
    \node at (-45:\r/2) {$#3$};
\end{tikzpicture}
}}}

\title{Extracting topological spins from bulk multipartite entanglement}
\author{Yarden Sheffer}
\affiliation{Department of Condensed Matter Physics, Weizmann Institute of Science Rehovot 7610001, Israel}
\author{Ady Stern}
\affiliation{Department of Condensed Matter Physics, Weizmann Institute of Science Rehovot 7610001, Israel}
\author{Erez Berg}
\affiliation{Department of Condensed Matter Physics, Weizmann Institute of Science Rehovot 7610001, Israel}
\begin{abstract}
We address the problem of identifying a 2+1d topologically ordered phase using measurements on the ground-state wavefunction. For non-chiral topological order, we describe a series of bulk multipartite entanglement measures that extract the invariants $\sum_a d_a^2 \theta_a^r$ for any $r \geq 2$, where $d_a$ and $\theta_a$ are the quantum dimension and topological spin of an anyon $a$, respectively. These invariants are obtained as expectation values of permutation operators between $2r$ replicas of the wavefunction, applying different permutations on four distinct regions of the plane. Our proposed measures provide a refined tool for distinguishing topological phases, capturing information beyond conventional entanglement measures such as the topological entanglement entropy. We argue that any operator capable of extracting the above invariants must act on at least $2r$ replicas, making our procedure optimal in terms of the required number of replicas. We discuss the generalization of our results to chiral states.
\end{abstract}
\date{\today}
\maketitle
Topologically ordered quantum states are a class of gapped quantum systems exhibiting unique properties, including anyonic excitations and robust ground-state degeneracy. Equivalently, such states can be defined as states that are not adiabatically connected to ``trivial" insulators, even in the absence of any protecting symmetries \cite{chen2010local}. In the absence of symmetries, the properties of the state are encoded in the entanglement pattern of the ground-state wavefunction \cite{Kitaev_Preskill_2006, Levin_Wen_2006}. In fact, it is now believed that all universal properties of the topological phase can be defined from the ground state \cite{Haah_2016, Kawagoe_Levin_2020, Shi_Kato_Kim_2020, Cian_Hafezi_Barkeshli_2022}. Such properties include the anyon dimensions $d_a$, the fusion rules, and data such as the modular $S$ and $T$ matrices, which, in most cases, suffice to identify the theory uniquely. 

While the ground-state wavefunction encodes, in principle, all information about the state, it remains an open challenge to define a set of measurements, in the bulk, that fully identifies the underlying state. The issue is that the topological phase should be invariant under finite-depth local unitary gates, and any probe of the phase should be robust to such perturbations. The problem of ``measuring" the state has gained attention recently in the context of quantum simulations of topologically-ordered states, where, for example, the topological Rényi entropy was used to demonstrate an underlying topologically-ordered state \cite{satzinger2021realizing}. Recent work derived explicit formulas that enable certain topological invariants to be calculated from the ground-state wavefunction alone. Examples include the chiral central charge \cite{tu2013momentum,Kim_Shi_Kato_Albert_2022}, the quantum hall conductance \cite{Fan_2023_extracting} or the many-body Chern number \cite{Cian_Hafezi_Barkeshli_2022}, the anyon quantum dimensions \cite{Liu_2024}, and the higher central charge \cite{kobayashi_extracting_2024}. However, these invariants generally do not fully characterize the phase, or require specific geometry and translation symmetry (in the case of \cite{tu2013momentum,kobayashi_extracting_2024}). Identifying a sufficient set of bulk measurements remains an open challenge.

In this work, we introduce an operator acting on $2r$ ($r\ge2$) replicas of the wavefunction $\ket\psi$ that extracts the invariant 
\begin{equation}
    \Phi(r)=\frac{1}{D}\sum_a \theta_a^r d_a^2
\end{equation} 
of a (non-chiral) topologically ordered ground-state, where $d_a$ and $\theta_a$ are the quantum dimension and the topological spin of the anyon $a$, and $D^2=\sum_ad_a^2$. For abelian models, $\Phi(r)$ for different $r$ fully determine the $T$ matrix. As an example, $\Phi(2)$ distinguishes between the toric code and the double semion models \cite{levin_string-net_2005}, which have the same fusion rules and total quantum dimension, but anyonic excitations with different mutual statistics. In fact, for most cases of interest (including non-abelian models), these invariants determine the underlying state uniquely (as we elaborate below).

\begin{figure}
  \begin{center}
  \begin{tikzpicture}
    \def\r{2.1};
      \draw[draw=darkred,very thick] (0,0) circle (\r);
      \fill[fill=orange,fill opacity=.05] (-3.3,-2.4) rectangle (3.3,2.4);
      \draw[very thick,darkred] (0,0) -- (30:\r);
      \draw[very thick,darkred] (0,0) -- (150:\r);
      \draw[very thick,darkred] (0,0) -- (-90:\r);
      \node at (-1,1.5) {\LARGE \textbf{$A$}};
      \node at (-2.5,2) {\LARGE \textbf{$\Lambda$}};
      \node at (-\r/2,-\r*.6) {\LARGE \textbf{$B$}};
      \node at (+\r/2,-\r*.6) {\LARGE \textbf{$C$}};
      \begin{scope}[scale=.4,shift={(-1.5,2.0)}]
        \node at (1.5,2.2) {\scriptsize \textcolor{darkblue}{$\per{\a}$}};
          \foreach \x in {0,...,3}{
            \filldraw[darkblue] (\x,0) circle (.1);
            \filldraw[darkblue] (\x,1) circle (.1);
          }
          \foreach \x in {0,...,2}{
            \draw[thick,->,lightblue] (\x+.2,0) -- ++ (.6,0);
            \draw[thick,->,lightblue] (\x+.8,1) -- ++ (-.6,0);
          }
          \draw[thick,->,lightblue] (0,1.2) to[in=150,out=30] (3,1.2);
          \draw[thick,->,lightblue] (3,-0.2) to[in=-30,out=-150] (0,-.2);
      \end{scope}
      \begin{scope}[scale=.4,shift={(-4.0,-2.0)}]
        \node at (1.5,1.6) {\scriptsize \textcolor{darkblue}{$\per{\b}$}};
          \foreach \x in {0,...,3}{
            \filldraw[darkblue] (\x,0) circle (.1);
            \filldraw[darkblue] (\x,1) circle (.1);
            \draw[thick,<->,lightblue] (\x,0.2) -- ++ (0,0.6);
          }
      \end{scope}
      \begin{scope}[scale=.4,shift={(1.0,-2.0)}]
        \node at (1.5,1.6) {\scriptsize \textcolor{darkblue}{$\per{\g}$}};
          \foreach \x in {0,...,3}{
            \filldraw[darkblue] (\x,0) circle (.1);
            \filldraw[darkblue] (\x,1) circle (.1);
          }
          \foreach \x in {1,2,3}
            `\draw[thick,<->,lightblue] (\x-.2, .8) -- ++ (-.6,-.6);
            \draw[thick,->,lightblue,dotted] (-.6,.4) -- (-.2,1-.2);
            \draw[thick,->,lightblue,dotted] (3.6,.6) -- (3.2,.2);
      \end{scope}
  \end{tikzpicture}
  \end{center}
  \caption{The permutation operators $\per{\a},\per{\b},\per{\g}$ acting on the three regions. Presented here for $r=4$.}\label{fig:pizza}
\end{figure}
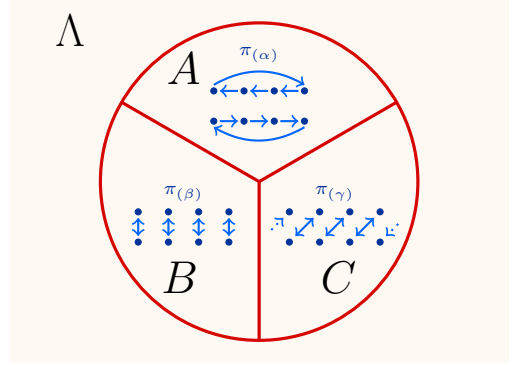

To construct the operator, we consider a region on the plane, partitioned into three parts $A,B,C$, as seen in Fig. \ref{fig:pizza} (with the external region denoted by $\Lambda$). The operator is defined by acting on three regions with three different permutations of the microscopic degrees of freedom on the $2r$ replicas (see App. \ref{app: perms} for an explicit definition of the permutations). Labeling each replica by the pair $(s,t)$ with $s=1,2$, $t=1,...,r$, we define the permutations as 
\begin{equation}
  \begin{aligned}
    \per{\alpha}(1,t) &= (1,t-1); & \per{\alpha}(2,t) &= (2,t+1), \\
    \per{\beta}(s,t) &= (s+1,t) \\ 
    \per{\gamma}(1,t) &= (2,t-1); & \per{\gamma}(2,t) &= (1,t+1).
  \end{aligned}
\end{equation}
Our main claim is that, for a non-chiral bosonic topologically-ordered state, in the limit where the length scale $L$ of all subregions is much larger than the correlation length, we have the following formula
\begin{equation}
\begin{aligned}
  \fabc &\equiv\mel{\rpsi}{\pi_{(\a)A}\pi_{(\b)B}\pi_{(\g)C}}{\rpsi}=Ce^{-2(r-1)L/\xi}\Phi(r),
  \label{eq: phi-def}
\end{aligned}
\end{equation}
where $C,\xi$ are real and non-universal. The term $C$ captures corner contributions to the multi-entropy, which were argued to have a universal lower bound \cite{siva2022universal, Liu_Kusuki_Kudler-Flam_Sohal_Ryu_2024, liu_multi_2024} for phases with ungappable boundaries, but can vary inside the phase. The scale $\xi$ is expected to behave as $a^2/l$ where $l$ is the correlation length and $a$ is a microscopic length scale. We argue that, while the multiplicative terms are non-universal, the phase of \eqref{eq: phi-def} is universal, and the magnitude can be obtained by appropriately canceling the non-universal contributions. This enables extracting $\Phi(r)$ universally.

We note that definitions related to \eqref{eq: phi-def} were considered under the term ``multi-entropy", where the Rényi entropy is generalized by considering different permutation operators applied on different regions, thereby obtaining a multipartite entanglement measure. Such measures were suggested in the context of the AdS/CFT correspondence \cite{Gadde_Krishna_Sharma_2022, penington_fun_2023}, and more recently on 2+1d topologically ordered states \cite{liu_multi_2024}. Importantly, these measures are related only to multipartite entanglement between the regions $A,B,C,\Lambda$, as they are invariant under local operations and classical communications (LOCC) in the four regions.

We note that a similar quantity to \eqref{eq: phi-def} (namely, the higher central charge $\zeta_r=\frac{\Phi(r)}{\abs{\Phi(r)}}$ \cite{ng2019higher,kaidi2022higher}) was obtained in \cite{kobayashi_extracting_2024} from a single replica as the expectation value of applying a $\frac{2\pi}{r}$ twist on one side of a cylinder. The result of \cite{kobayashi_extracting_2024} is obtained under the assumption of a CFT entanglement spectrum. The procedure we present is applied in the bulk (on an arbitrary manifold) and can be employed for systems with a gapped entanglement spectrum. Below, we also argue that, in the absence of symmetries, no operator acting on a single replica can extract the higher central charge. 

In the following, we give two arguments for the validity of \eqref{eq: phi-def}. The first is given using a topological quantum field theory (TQFT) approach, in a spirit similar to \cite{dong_topological_2008}. The second is by explicit calculation on exactly-solvable string-net models \cite{levin_string-net_2005}. The lattice calculation is also used to argue that the number of replicas $2r$ is optimal for an operator that extracts $\Phi(r)$. Finally, we discuss some consequences of our results, including some cases in which $\Phi(r)$ for different $r$ is sufficient to identify the underlying state uniquely.

\paragraph*{Preliminaries.}
Before proving \eqref{eq: phi-def}, we give a few notes on our construction. First, we have
\begin{equation}
  \begin{aligned}
    \per{\beta}\per{\gamma}&=\per{\alpha};&  \per{\beta}^2&=\per{\gamma}^2=1; & \per{\alpha}^r&=1.
  \end{aligned}
\end{equation}

\newcommand{\pert}[1]{\tilde{\pi}_{(#1)}}
Second, since $\ket{\rpsi}$ is invariant under any permutation of the replicas $\ket{\rpsi}=\sigma\ket{\rpsi}$, $\fabc$ is invariant under any conjugation of all three $\pi_i$, setting, $\tilde{\pi}_i =\sigma^\dagger \pi_i \sigma$. In particular, we can choose $\sigma$ such that $\pert{\b}=\per{\g},\pert{\g}=\per{\b},\pert{\a}=\per{\a}^\dagger$. As a result, we see that under spatial reflection which exchanges the regions $B\leftrightarrow C$ we have $\fabc\mapsto\fabc^*$. This agrees with the expected behavior of \eqref{eq: phi-def} since under reflection we have $\theta_a\mapsto \theta_a^*$.

Third, we will show the following relation, in which the nonuniversal terms in \eqref{eq: phi-def} cancel:
\begin{equation}
    \frac{1}{D^{r}}\abs{\Phi(r)}^2=\frac{\pizza{\a}{\b}{\g}\ \pizza{\g}{\b}{\a}\ \pizza{\a}{}{\a}\ \pizza{\g}{}{\g}}{\pizza{\a}{}{\g}\ \pizza{\g}{}{\a}\ \pizza{\g}{\b}{\g}\ \pizza{\a}{\b}{\a}},
    \label{eq: phi-mag}
\end{equation}
where $\tpizza{\a}{\b}{\g}=\fabc$ etc. (an absent symbol means that no permutation is applied in that region). Importantly, the RHS of \eqref{eq: phi-mag} is invariant under local perturbations, as well as under local modifications of the boundaries, as the edge and corner terms cancel between the numerator and the denominator. This is clearly seen for all external edges and corners. For the internal corner, this can be seen by applying $\per{\b}$ on all regions in the denominator (at the cost of introducing a permutation in $\Lambda$ as well). This also explains why the phase in \eqref{eq: phi-def} is universal: any change in $\fabc$ from a local unitary on the boundary of the regions can be canceled by one of the denominator terms. For example, the term $\tpizza{\a}{\b}{\g}/\tpizza{\a}{}{\g}$ is invariant under changes in the upper-right corner. However, the denominator terms are all real, so the phase is unchanged by local unitaries.


Note that some fine-tuned examples exist in which the topological contribution to the entanglement entropy \cite{Zou_Haah_2016, williamson2016spurious} or the modular commutator \cite{Gass_Levin_2024} attains a non-universal value. Nevertheless, it is believed that for a ``generic" state in the phase the results are universal \cite{Kitaev_Preskill_2006, Levin_Wen_2006, Oliviero_Leone_Zhou_Hamma_2022}. The same caveats apply in our case.

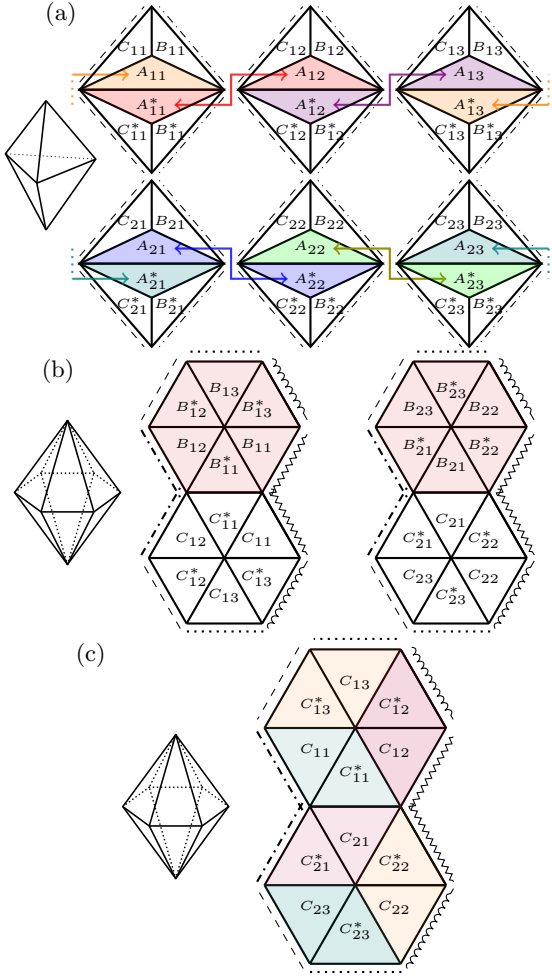
\begin{figure}
  \begin{center}
  \newcommand{\sorigami}[4]{
  \def\r{1.1};
  \begin{scope}[every path/.append style={thick},every node/.append style={font={\tiny}}]
  \begin{scope}[shift={(0,\r/2)}]
  \coordinate (a) at (0,\r*.5);
  \coordinate (b) at (-{sqrt(3)/2*\r},-\r/2);
  \coordinate (c) at ({sqrt(3)/2*\r},-\r/2);
  \coordinate (d) at (0,-.1);
  \draw (a) -- (b) -- (c) -- cycle;
  \draw (d) -- (a);
  \draw (d) -- (b);
  \draw (d) -- (c);
  \fill[#3] (c) -- (b) -- (d) -- cycle;
  \node at (-90+0*120:\r*.3) {$A_{#1#2}$};
  \node at (.25,.00) {$B_{#1#2}$};
  \node at (-.25,.00) {$C_{#1#2}$};
  \end{scope}
  \begin{scope}[shift={(0,-\r/2)}]
  \coordinate (a) at (0,-\r*.5);
  \coordinate (b) at (-{sqrt(3)/2*\r},\r/2);
  \coordinate (c) at ({sqrt(3)/2*\r},\r/2);
  \coordinate (d) at (0,.1);
  \draw (a) -- (b) -- (c) -- cycle;
  \draw (d) -- (a);
  \draw (d) -- (b);
  \draw (d) -- (c);
  \fill[#4] (c) -- (b) -- (d) -- cycle;
  \node at (90-0*120:\r*.3) {$A^*_{#1#2}$};
  \node at (.25,.00) {$B^*_{#1#2}$};
  \node at (-.25,.00) {$C^*_{#1#2}$};
  \end{scope}
  \draw[thin, dashed] ({-sqrt(3)/2*\r-.0},.1) -- (-.1,\r);
  \draw[thin, dash dot] ({sqrt(3)/2*\r-.0},.1) -- (.1,\r);
  \draw[thin, dashed] ({-sqrt(3)/2*\r-.0},-.1) -- (-.1,-\r);
  \draw[thin, dash dot] ({sqrt(3)/2*\r-.0},-.1) -- (.1,-\r);
  \end{scope}
  }
  
  \begin{tikzpicture}
  \node at (-1.2,1) {(a)};
  \def\s{1.1}
  \def\dif{2.1}
  \node at (-1.5,-1) {
  \tdplotsetmaincoords{70}{100}     
  \begin{tikzpicture}[tdplot_main_coords,scale=.7]
      \coordinate (A) at (xyz cylindrical cs:z=1.3);
      \coordinate (B) at (xyz cylindrical cs:z=-1.3);
      \foreach \t in {0,120,240}{
        \draw[semithick] (A) -- (xyz cylindrical cs: radius=1,angle=\t) -- (B);
      }
    \draw[semithick] (xyz cylindrical cs: radius=1,angle=0) -- (xyz cylindrical cs: radius=1, angle=120);
    \draw[dots=20 per 1cm] (xyz cylindrical cs: radius=1,angle=120) -- (xyz cylindrical cs: radius=1, angle=240);
    \draw[semithick] (xyz cylindrical cs: radius=1,angle=240) -- (xyz cylindrical cs: radius=1, angle=360);
  \end{tikzpicture}
  };
  \begin{scope}[shift={(0,0)}]
      \draw[<->,red,opacity=.8,thick] (.3,-.2) -- (\dif/2,-.2) -- (\dif/2,.2) -- (\dif-.3,.2);
      \draw[dotted,orange,opacity=.8,thick] (-\dif/2,-.2) -- (-\dif/2,.2);
      \draw[->,orange,opacity=.8,thick] (-\dif/2,.2) -- (-.3,.2);
      \sorigami{1}{1}{orange,opacity=.2}{red,opacity=.2}
  \end{scope}
  \begin{scope}[shift={(\dif,0)}]
      \draw[<->,violet,opacity=.8,thick] (.3,-.2) -- (\dif/2,-.2) -- (\dif/2,.2) -- (\dif-.3,.2);
      \sorigami{1}{2}{red,opacity=.2}{violet,opacity=.2}
  \end{scope}
  \begin{scope}[shift={(2*\dif,0)}]
      \sorigami{1}{3}{violet,opacity=.2}{orange,opacity=.2}
      \draw[dotted,orange,opacity=.8,thick] (\dif/2,-.2) -- (\dif/2,.2);
      \draw[->,orange,opacity=.8,thick] (\dif/2,-.2) -- (.3,-.2);
  \end{scope}
  \begin{scope}[shift={(0,-2.3)}]
  \begin{scope}[shift={(0,0)}]
      \sorigami{2}{1}{blue,opacity=.2}{teal,opacity=.2}
      \draw[<->,blue,opacity=.8,thick] (.3,.2) -- (\dif/2,.2) -- (\dif/2,-.2) -- (\dif-.3,-.2);
      \draw[dotted,teal,opacity=.8,thick] (-\dif/2,.2) -- (-\dif/2,-.2);
      \draw[->,teal,opacity=.8,thick] (-\dif/2,-.2) -- (-.3,-.2);
  \end{scope}
  \begin{scope}[shift={(\dif,0)}]
      \sorigami{2}{2}{green,opacity=.2}{blue,opacity=.2}
      \draw[<->,olive,opacity=1,thick] (.3,.2) -- (\dif/2,.2) -- (\dif/2,-.2) -- (\dif-.3,-.2);
  \end{scope}
  \begin{scope}[shift={(2*\dif,0)}]
      \sorigami{2}{3}{teal,opacity=.2}{green,opacity=.2}
      \draw[dotted,teal,opacity=.8,thick] (\dif/2,.2) -- (\dif/2,-.2);
      \draw[->,teal,opacity=.8,thick] (\dif/2,.2) -- (.3,.2);
  \end{scope}
  \end{scope}
  \end{tikzpicture}
  
  \begin{tikzpicture}[every node/.append style={font={\tiny}},every path/.append style={thick}]
  \node at (-3.7,1.6) {\small (b)};
  \node at (-3.7,0) {
  \tdplotsetmaincoords{65}{120}
  \begin{tikzpicture}[tdplot_main_coords,scale=.7]
      \coordinate (A) at (xyz cylindrical cs:z=1.5);
      \coordinate (B) at (xyz cylindrical cs:z=-1.5);
      \foreach \t in {180,240}{
        \draw[semithick,dots=20 per 1cm] (A) -- (xyz cylindrical cs: radius=1,angle=\t) -- (B);
      }
      \foreach \t in {300,360,60,120}{
        \draw[semithick] (A) -- (xyz cylindrical cs: radius=1,angle=\t) -- (B);
      }
      \foreach \t in {120,180,240}
        \draw[semithick,dots=20 per 1cm] (xyz cylindrical cs: radius=1,angle=\t) -- (xyz cylindrical cs: radius=1,angle=60+\t);
      \foreach \t in {300,0,60}
        \draw[semithick] (xyz cylindrical cs: radius=1,angle=\t) -- (xyz cylindrical cs: radius=1,angle=60+\t);
  \end{tikzpicture}
  };
  \def\r{1.0}
  \def\rr{1.1}
  \begin{scope}[shift={(-1.5,0)}]
  \origami{\r}{\rr};
  \begin{scope}[shift={(0,0.866025404*\r)}]
   \filldraw[darkred,opacity=.1] (0:\r) \foreach \x in {60,120,...,360} {  -- (\x:\r) };
      \node at (-30:\r/2) {$B_{11}$};
      \node at (-30+1*60:\r/2) {$B^*_{13}$};
      \node at (-30+2*60:\r/2) {$B_{13}$};
      \node at (-30+3*60:\r/2) {$B^*_{12}$};
      \node at (-30+4*60:\r/2) {$B_{12}$};
      \node at (-30+5*60:\r/2) {$B^*_{11}$};
  \end{scope}
  \begin{scope}[shift={(0,-0.866025404*\r)}]
      \node at (30:\r/2) {$C_{11}$};
      \node at (30-1*60:\r/2) {$C^*_{13}$};
      \node at (30-2*60:\r/2) {$C_{13}$};
      \node at (30-3*60:\r/2) {$C^*_{12}$};
      \node at (30-4*60:\r/2) {$C_{12}$};
      \node at (30-5*60:\r/2) {$C^*_{11}$};
  \end{scope}
  \end{scope}
  \begin{scope}[shift={(1.5,0)}]
  \origami{\r}{\rr};
  \begin{scope}[shift={(0,0.866025404*\r)}]
   \filldraw[darkred,opacity=.1] (0:\r) \foreach \x in {60,120,...,360} {  -- (\x:\r) };
      \node at (-30:\r/2) {$B^*_{22}$};
      \node at (-30+1*60:\r/2) {$B_{22}$};
      \node at (-30+2*60:\r/2) {$B^*_{23}$};
      \node at (-30+3*60:\r/2) {$B_{23}$};
      \node at (-30+4*60:\r/2) {$B^*_{21}$};
      \node at (-30+5*60:\r/2) {$B_{21}$};
  \end{scope}
  \begin{scope}[shift={(0,-0.866025404*\r)}]
      \node at (30:\r/2) {$C^*_{22}$};
      \node at (30+-1*60:\r/2) {$C_{22}$};
      \node at (30+-2*60:\r/2) {$C^*_{23}$};
      \node at (30+-3*60:\r/2) {$C_{23}$};
      \node at (30+-4*60:\r/2) {$C^*_{21}$};
      \node at (30+-5*60:\r/2) {$C_{21}$};
  \end{scope}
  \end{scope}
  \end{tikzpicture}
  
  \begin{tikzpicture}[every node/.append style={font={\tiny}},every path/.append style={thick}]
  \node at (-3.5,2.0) {\small (c)};
  \node at (-2.5,0) {
  \tdplotsetmaincoords{65}{120}
  \begin{tikzpicture}[tdplot_main_coords,scale=.7]
      \coordinate (A) at (xyz cylindrical cs:z=1.5);
      \coordinate (B) at (xyz cylindrical cs:z=-1.5);
      \foreach \t in {180,240}{
        \draw[semithick,dots=20 per 1cm] (A) -- (xyz cylindrical cs: radius=1,angle=\t) -- (B);
      }
      \foreach \t in {300,360,60,120}{
        \draw[semithick] (A) -- (xyz cylindrical cs: radius=1,angle=\t) -- (B);
      }
      \foreach \t in {120,180,240}
        \draw[semithick,dots=20 per 1cm] (xyz cylindrical cs: radius=1,angle=\t) -- (xyz cylindrical cs: radius=1,angle=60+\t);
      \foreach \t in {300,0,60}
        \draw[semithick] (xyz cylindrical cs: radius=1,angle=\t) -- (xyz cylindrical cs: radius=1,angle=60+\t);
  \end{tikzpicture}
  };
  \def\r{1.2}
  \def\rr{1.3}
  \origami{\r}{\rr};
  \begin{scope}[shift={(0,0.866025404*\r)}]
   \filldraw[teal,opacity=.1] (0,0) -- (-60:\r) -- (-120:\r) -- (180:\r) -- cycle;
   \filldraw[orange,opacity=.1] (0,0) -- (-180:\r) -- (120:\r) -- (60:\r) -- cycle;
   \filldraw[purple,opacity=.15] (0,0) -- (60:\r) -- (0:\r) -- (-60:\r) -- cycle;
      \node at (-30:\r/2) {$C_{12}$};
      \node at (-30+1*60:\r/2) {$C^*_{12}$};
      \node at (-30+2*60:\r/2) {$C_{13}$};
      \node at (-30+3*60:\r/2) {$C^*_{13}$};
      \node at (-30+4*60:\r/2) {$C_{11}$};
      \node at (-30+5*60:\r/2) {$C^*_{11}$};
  \end{scope}
  \begin{scope}[shift={(0,-0.866025404*\r)}]
       \filldraw[orange,opacity=.1] (0,0) -- (60:\r) -- (0:\r) -- (-60:\r) -- cycle;
       \filldraw[purple,opacity=.1] (0,0) -- (-180:\r) -- (120:\r) -- (60:\r) -- cycle;
       \filldraw[teal,opacity=.15] (0,0) -- (180:\r) -- (240:\r) -- (-60:\r) -- cycle;
      \node at (30:\r/2) {$C^*_{22}$};
      \node at (30-1*60:\r/2) {$C_{22}$};
      \node at (30-2*60:\r/2) {$C^*_{23}$};
      \node at (30-3*60:\r/2) {$C_{23}$};
      \node at (30-4*60:\r/2) {$C^*_{21}$};
      \node at (30-5*60:\r/2) {$C_{21}$};
  \end{scope}
  \end{tikzpicture}
  \end{center}
  \caption{The gluing scheme used for the calculation of $\fabc$. We illustrate here with $r=3$ and the pyramids flattened so that the faces are shown. The pyramids are obtained by identifying the decorated edges, the corresponding 3d shapes are drawn on the left. (a) Gluing of the $A$ faces. The result is two double pyramids. (b) the $B$ faces of the double pyramids are glued together. (c) The gluing of the $C$ faces results in gluing the top and bottom of the double pyramid with a $2\pi/r$ twist, obtaining the lens space $L(r,1)$.}\label{fig: diamonds}
  
\end{figure}

\paragraph*{TQFT argument.}
We now calculate $\fabc$ based on a TQFT approach. This approach only gives the universal part of \eqref{eq: phi-def}, missing the prefactor $C$ (that includes corner contributions) and the area-law term, which come from short-distance scale physics. Again, note that these nonuniversal contributions cancel out in \eqref{eq: phi-mag}. The idea, similar to \cite{dong_topological_2008}, is to represent the expectation value of the operator as the partition function of the TQFT describing the phase on a manifold obtained by gluing the regions $A,B,C$ according to the permutations $\per{i}$. To begin, we consider the density matrix $\rho=\rm {Tr}_{\Lambda}\ket{\psi}\bra{\psi}$. In TQFT, $\rho$ is represented by a three-dimensional ball. Here we divide the surface of the ball into 6 faces labeled by $i_{s,t}$ or $i^*_{s,t}$ for the ket and bra parts in replica $s,t$, respectively, that is 
\begin{equation}
  \rho=D\cdot\vcenter{\hbox{
  \begin{tikzpicture}[every path/.append style={thick}, every node/.style={font={\tiny}}]
    \coordinate (a) at (0,0);
    \coordinate (b) at (1.4,.3);
    \coordinate (c) at (.8,-.2);
    \coordinate (d) at (.7,1.0);
    \coordinate (e) at (.7,-1.0);
    \draw[thin,->] (0,.5) to[out=100, in=120] (.6,.5);
    \draw[thin,->] (.1,-.8) to[out=-80, in=-120] (.6,-.6);
    \filldraw[fill=white,fill opacity=.65] (a) -- (d) -- (b) -- (e) -- cycle;
    \draw (a) -- (c) -- (b);
    \draw[dashed] (a) -- (b);
    \draw (d) -- (c) -- (e);
    \node at (.5,.2) {$A_{s,t}$};
    \node at (.5,-.4) {$A_{s,t}^*$};
    \node at (1.05,.35) {$B_{s,t}$};
    \node at (1.1,-.4) {$B_{s,t}^*$};
    \node at (.0,.4) {$C_{s,t}$};
    \node at (.0,-.7) {$C_{s,t}^*$};
  \end{tikzpicture}
  }}.
\end{equation}
The normalization of $\rho$ is obtained by requiring $\Tr \rho=1$. The trace is given by gluing the top and bottom of the ball, obtaining the partition function $Z(S^3)=D^{-1}$ \cite{dong_topological_2008}. We now glue together the balls to obtain the partition function that corresponds to $\fabc$. 

\begin{figure}
  \begin{center}
    \includegraphics[width=0.35\textwidth]{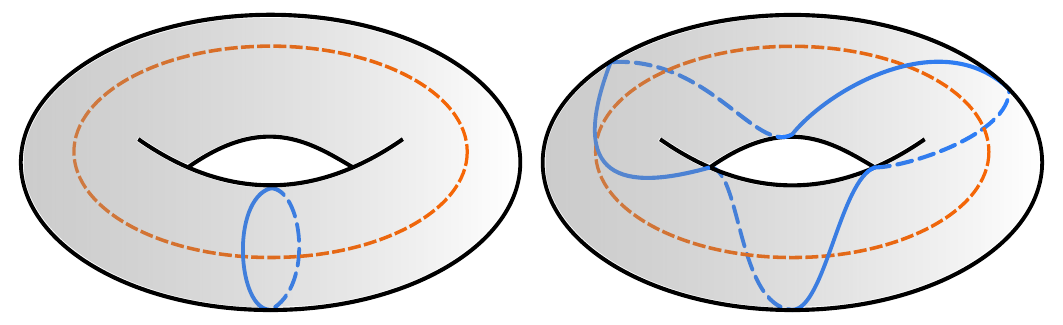}
  \end{center}
  \caption{Obtaining the lens space $L(r,1)$ by applying a Dehn twist on a solid torus along its longitude (orange dashed line), then gluing it to an untwisted solid torus. The action of the Dehn twist on the vacuum sector gives the partition function \eqref{eq: lens-partition-function}.}\label{fig:tori}
\end{figure}

We begin by gluing the $A$-faces, gluing $A_{(s,t)}$ to $A^*_{\per{\alpha}(s,t)}$ (Fig. \ref{fig: diamonds}a). This gives two double pyramids with $2r$ edges on their base, corresponding to $s=1,2$, whose $t$ indices go in opposite directions (Fig. \ref{fig: diamonds}b). Next, we glue the $B$ faces, connecting the two double pyramids to a single double pyramid (with the topology of a 3-ball). Finally, we consider the action of the gluing of the $C$ faces (Fig. \ref{fig: diamonds}c). This glues the top and bottom pyramids with a $\frac{2\pi}{r}$ twist. Mathematically, the resulting space 3-manifold is the lens space $L(r,1)$. We therefore find that 
\begin{equation}
  \fabc^{\mathrm{TQFT}} = D^{2r} Z(L(r,1)).
  \label{eq: phi-from-lens}
\end{equation}
The partition function $Z(L(r,1))$ is known \cite{freed1991computer,kaidi2022higher}, but is derived here for completeness. An alternative definition of $L(r,1)$ is obtained by taking the solid torus, applying $r$ Dehn twist operations along its longitudinal axis (the axis that goes through the torus), then gluing its meridian (the line that goes around the handle) to a second solid torus \cite{rolfsen2003knots} (see Fig. \ref{fig:tori}). The partition function can then be obtained from the appropriate application of modular transformations, noting that both solid tori are in the vacuum sector. This gives
\begin{equation}
  Z(L(r,1))=(S^\dagger T^rS)_{00} =\frac{1}{D^2} \sum_a d_a^2 \theta_a^r,
  \label{eq: lens-partition-function}
\end{equation}
where $S,T$ are the modular matrices. Here we used the fact that the phase is non-chiral. Otherwise, one obtains a phase contribution arising from the ``framing anomaly" when $c_-\neq 0$ \cite{witten1989quantum, Witten_1990}. 

A similar calculation can be used to obtain the other terms in \eqref{eq: phi-mag}, noticing that the only manifolds that are obtained are the 3-sphere, giving a $D^{-1}$ contribution for each sphere, and the mirror image of the lens space $L(r,-1)$. As explained above, the result for equation \eqref{eq: phi-mag} is universal.

\paragraph*{Lattice model derivation.} It is useful to verify the TQFT argument by an explicit calculation on the lattice. We now calculate $\fabc$ for a string-net model \cite{levin_string-net_2005}. Similar to the case of TEE, the calculation should be valid for generic wavefunctions in the same phase. Below we describe the calculation in the simplest case of a twisted $\Z_N$ gauge theory (equivalently, a string-net model with $\Z_N$ strings) \cite{Hu_Wan_Wu_2013_twisted,lin_generalizations_2014}. It is also possible to carry the calculation for more general string-net models. We describe the general scheme in App. \ref{app: string-net}. 

\begin{figure}
\begin{tikzpicture}
\def\r{2.2};
\begin{scope}[every path/.append style={very thick,darkred},yscale=.7]
      \draw (0,0) circle (\r);
      \draw (0,0) -- (30:\r);
      \draw (0,0) -- (150:\r);
      \draw (0,0) -- (-90:\r);
\end{scope}
\coordinate (O) at (0,2);
\coordinate (A) at (0,.7);
\coordinate (B) at (-1.5,-.5);
\coordinate (C) at (1.5,-.5);
\begin{scope}[every path/.append style={thick,lightblue}]
    \draw[midarr] (O) --node[right] {$a$} (A);
    \draw[midarr] (B) --node[left=.2,pos=.6] {$b$} (A);
    \draw[midarr] (B) --node[above,pos=.6] {$c$} (C);
    \draw[midarr] (A) --node[right=.2,pos=.4] {$a+b$} (C);
    \draw[midarr] (O) to[in=135,out=190,looseness=2] node[above=.3] {$b+c$} (B);
    \draw[midarr] (C) to[out=45,in=-10,looseness=2] node[above=.30] {$a+b+c$} (O);
\end{scope}
\node at (1,-1) {\large $C$};
\node at (-1,-1) {\large $B$};
\node at (-.7,1) {\large $A$};
\end{tikzpicture}
  \caption{The anyonic charges passing between the regions. The permutations $\pi_{A,B,C}$ exchange the vertices, and we consider only configurations of the anyonic charges such that the charge lines are not ``cut" by the action of the permutation.}\label{fig: reduced-tetrahedron}
\end{figure}
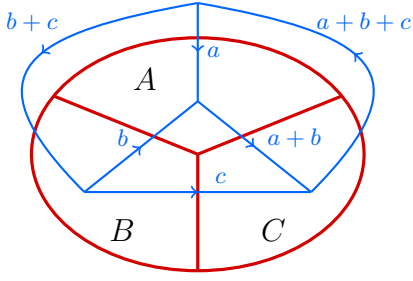

Since our scheme cancels out area-law contributions, we can calculate $\fabc$ on the simplest string-net graph, where qudits are defined on the edges of a tetrahedron, as in Fig. \ref{fig: reduced-tetrahedron}.  In App. \ref{app: string-net} we provide a full calculation for a general string-net model, and show that the area-law term follows the behavior of Eq. \eqref{eq: phi-def}, and is canceled in the expression \eqref{eq: phi-mag}, as expected.

In the model describing twisted $\Z_N$ gauge theory, each edge carries a ``string" labeled by numbers $a=0,...,N-1$, with the requirement that when strings $a,b,c$ meet at a vertex we have $a+b+c=0\mod N$. The state on the tetrahedron is specified uniquely by the choice of $a,b,c$ in Fig. \ref{fig: reduced-tetrahedron}. The ground state on the tetrahedral graph is then given by \cite{levin_string-net_2005}
\begin{equation}
    \ket\y =\frac{1}{N^{3/2}}\sum_{abc}F(a,b,c)\ket{a,b,c},
    \label{eq: psi-string-net}
\end{equation}
where $F(a,b,c)$ are the group cohomology factors, given by \cite{lin_generalizations_2014}
\begin{equation}
  F(a,b,c)=\exp(\frac{2\pi i p}{N}a\qty{b,c}),
  \label{eq:Z_N-f-sym}
\end{equation}
where $\qty{b,c}=1$ if $b+c\ge N$ and 0 otherwise.  The integer $p=0,...,N-1$ distinguishes the $N$ different ``twists" of the $\Z_N$ gauge theory, which differ by the topological spins of the anyons. The case of $p=0$ reduces to the $\Z_N$ toric code \cite{Kitaev_2003}. For $N=2$, the case $p=1$ is the double semion model \cite{levin_string-net_2005}.

The calculation of $\fabc$ reduces to the summation over configurations of the $2r$ tetrahedra such that the anyon lines on the edges are not ``cut" by the permutations \footnote{formally, one should think of the edge as having two maximally entangled qubits, one on each side of the boundary between the regions}. That is, if there is an anyon line $a_{(s,t)}$ passing between regions $i,j$ on replica $(s,t)$, we must have
\begin{equation}
    a_{\pi_i^{-1}(s,t)}=a_{\pi_j^{-1}(s,t)}.
\end{equation}
This significantly constrains the tetrahedra that have to be summed over. Explicitly, we obtain
\begin{equation}
  \begin{aligned}
    a_{s,t}&=a_{s,t'}\equiv A_s \\ 
    b_{1,t}&=b_{2,t-1}\equiv B_t \\ 
    c_{s,t}&=c_{s,t'}\equiv C_s \\ 
    (a+b)_{1,t}&=(a+b)_{2,t-2}\Rightarrow &A_2-A_1&=B_t-B_{t-1} \\
    (b+c)_{1,t}&=(b+c)_{2,t}\Rightarrow &C_2-C_1&=A_2-A_1.
  \end{aligned}
  \label{eq: Z_N-constraints}
\end{equation}
\newcommand{\td}{\tilde{\delta}}
We see that the allowed $a,b,c$ depend only on $A_1,B_1,C_1,\td\equiv A_2-A_1$. We set $A=A_1,B=B_1,C=C_1$, such that the sum is carried only over $A,B,C,\td$. 

Before carrying out the sum, we note some features of the resulting constraints. Notice that, for $\td=0$, the $2r$ tetrahedra all have the same configuration. In this case, the phase factors $F$ are canceled between the bra and ket. The configurations that give non-trivial ($p$ dependent) contributions are, therefore, only those with $\td\neq 0$. In addition, $\td$ has to satisfy $r\td=0$ (mod $N$). This means that the sum includes non-trivial contributions only when $g=\gcd(r,N)>1$ (gcd is the greatest common divisor of $r,N$). In that case, we sum over $\td=\frac{N}{g}\delta$ and $\delta=1,...,g$. For simplicity, we assume below that $r=N$ and $N$ is prime, so $\td = \delta$. The derivation for general $r$ and $N$ is given in App. \ref{app: abelian general r}.

Now for the calculation. From \eqref{eq: psi-string-net} and \eqref{eq: Z_N-constraints}, $\fabc$ becomes
\begin{widetext}
\begin{equation}
  \begin{aligned}
    \fabc&=\frac{1}{N^{6r}}\sum_{A,B,C,\delta} \prod_{t=1}^rF(A,B+\delta,C)F(A+\delta,B+(t+1)\delta,C+\delta)F(A,B+(t+1)\delta,C+\delta)^* F(A+\delta,B+t\delta,C)^* \\ 
            &=\frac{1}{N^{6r-1}}\sum_{B,C,\delta} \exp\qty{\frac{2\pi i p\delta}{N} \sum_t\qty[\qty{B+t,C+\delta}-\qty{B+t,C}]}.
\end{aligned}
  \label{eq: phi-abelian}
\end{equation}
\end{widetext}
The sum inside the exponent is independent of $B$ and $C$. We get
\begin{equation}
  \fabc = \frac{1}{N^{6r-3}}\sum_\delta \exp\qty(\frac{2\pi i p\delta^2}{N})=\frac{1}{N^{6r-2}}\Phi(r).
  \label{eq: phi-abelian-result}
\end{equation}
The last equality is obtained by explicit calculation based on the $K$-matrix, see App. \ref{app: Zn twisted gauge}. Collecting the additional terms in \eqref{eq: phi-mag} gives the desired result.

\paragraph*{Optimality of the operator.}
\newcommand{\ok}{{\otimes k}}
We have shown above that $\Phi(r)$ can be extracted from the expectation values of operators acting on $2r$ replicas. In particular, the phase of $\Phi(r)$ can be extracted from a single such operator (see Eq. \eqref{eq: phi-def}). We now argue that this phase cannot be extracted from the expectation value of an operator acting on less than $2r$ replicas. To show that, we argue that the expectation value of any operator acting on less than $2N$ replicas cannot distinguish between two different twisted $\Z_N$ gauge theories. Since the phase of $\Phi(N)$ can differ between the two theories, it cannot be extracted using such an operator. This is formalized in the following:
\begin{conjecture}[1]\label{thm: optimality}
  Let $\ket{\psi_{1,2}}$ be two string-net wavefunctions corresponding to two different twisted $\Z_N$ gauge theories (different $p$'s), with $N$ prime. There exists a family of shallow random circuits $U$, such that, for any operator $\mathcal{O}$ acting on the space of $k<2N$ replicas, we have 
\begin{equation}
  \begin{aligned}
  &\underset{U}{\mathbb{E}}\mel{\psi_1^\ok}{\qty(U^\ok)^\dagger \mathcal{O}U^\ok}{\psi^\ok_1}\\
  =&\underset{U}{\mathbb{E}}\mel{\psi_2^\ok}{\qty(U^\ok)^\dagger \mathcal{O}U^\ok}{\psi^\ok_2},
\end{aligned}
  \label{eq: fooling-o}
  \end{equation}
where $\mathbb{E}_U$ denotes the expectation value over realizations of $U$.
\end{conjecture}
 
The above unitaries can be written as a product of local phase gates on each vertex
\begin{equation}
  U=\prod_v U_v,
  \label{eq: u-vertex-decomposition}
\end{equation}
where
\begin{equation}
  \begin{aligned}
    U_v\ket{
      \vcenter{\hbox{
      \begin{tikzpicture}[every path/.style={thick}, every node/.style={font={\scriptsize}},scale=.7]
        \draw[thick,midarr] (-30:1) --node[above] {$a$} (0,0);
        \draw[thick,midarr] (-30+240:1) --node[above] {$b$} (0,0);
        \draw[thick,midarr] (0,0) --node[right] {$a+b$} (0,1);
  \end{tikzpicture}}}
      }&=e^{i \phi_{ab}}\ket{
      \vcenter{\hbox{
      \begin{tikzpicture}[every path/.style={thick}, every node/.style={font={\scriptsize}},scale=.7]
        \draw[thick,midarr] (-30:1) --node[above] {$a$} (0,0);
        \draw[thick,midarr] (-30+240:1) --node[above] {$b$} (0,0);
        \draw[thick,midarr] (0,0) --node[right] {$a+b$} (0,1);
  \end{tikzpicture}}}
};& \phi_{ab}&\in\qty[0,2\pi].
  \end{aligned}
  \label{eq: uv-req}
\end{equation}
The phases $\phi_{ab}$ on each vertex are chosen uniformly and independently at random. Thus, each random $U$ gives a phase that depends on the string configuration. Intuitively, the claim means that there is no operator acting on fewer than $2N$ replicas that can distinguish between the topological phases obtained from $\ket{\psi_1}$ and $\ket{\psi_2}$. We support the claim by proving it in App. \ref{app: operator optimality proof} when the edge lattice is in the reduced form of Fig. \ref{fig: reduced-tetrahedron}. The argument can be summarized as follows: The operators $U$ are local phase gates, which can locally transform the $F$ symbols. The minimal configurations whose phase is invariant under these transformations are exactly those that are summed over in \eqref{eq: Z_N-constraints}. Notice that the operators $U$ applied on the wavefunction will not change the value of $\fabc$. 

We note that the unitaries $U$ explicitly break translation invariance. When translation invariance is imposed, it is possible to obtain universal information using fewer replicas (as the examples of \cite{tu2013momentum,kobayashi_extracting_2024} show). In App. \ref{app: operator optimality proof} we provide a simple argument showing that, in the absence of symmetries, no information about the topological order can be obtained from a single replica.

\paragraph*{Discussion.}
While in this work we described measurements on multiple replicas of the system, in practice these could be substituted by multiple randomized measurements on a single replica \cite{elben2023randomized, huang2020predicting,elben2023randomized}. This does not contradict the optimality discussed above, as it requires the expectation values of a large number of operators. Note, however, that the required number of such measurements will be exponential in $2rL$ where $L$ is the size of the regions.

Our methods are not sufficient to calculate $\fabc$ for phases with non-zero chiral central charge. It remains an open problem to complete the calculation in that case, for example, using boundary CFT methods \cite{Liu_Kusuki_Kudler-Flam_Sohal_Ryu_2024}. We note that \eqref{eq: phi-mag} remains true for the chiral case, but a phase contribution of the form $e^{if(r)c_-}$ should be added to \eqref{eq: phi-def}.

An interesting question about our result \eqref{eq: phi-def} is whether the knowledge of $\Phi(r)$ for different $r$ is sufficient to identify the phase of the underlying state uniquely. In the case of abelian topological order, the values $\Phi(r)$, together with the topological entanglement entropy, give $\Tr{T^r}$ for any $r$ and thus allow one to obtain the entire list of spins $\theta_a$. In all examples of abelian theories we are familiar with, the list of spins is sufficient to determine the state uniquely, but it is interesting to ask whether it is true in general (formally, both the spins and the fusion rules are required to fully specify the topological order). For non-abelian states, the topological order can be uniquely determined for theories with a small enough number of anyons, e.g. those classified in \cite{Ng_Rowell_Wang_Wen_2023}, but cannot be determined in general \cite{Mignard_Schauenburg_2021}. It is an interesting open problem to extract topological invariants that can distinguish any two phases, even beyond modular data \cite{Bonderson_2018_beyond}, from bulk wavefunctions.

\paragraph*{Acknowledgments.}
We thank Xiang Li, Bowen Shi, Noa Feldman, Evgenii Zheltonozhskii, Shinsei Ryu, and Roman Geiko for useful discussions. EB and AS thank the Kavli Institute of Theoretical Physics for their hospitality. This work was supported by grants from the DFG (CRC/Transregio 183, EI 519/71, projects C02 and C03), and the ISF Quantum Science and Technology program (Grant no. 2478/24). YS is supported by the Chaim Mida Prize for Outstanding Students in Theoretical Physics.

\bibliography{bibliography.bib}
\onecolumngrid
\appendix
\section{Explicit definition of the permutation operator for a bosonic system}
\label{app: perms}
Here we define the permutation operator acting in a region $A$ for a bosonic system. The operators discussed in the main text are obtained as products of permutation operators in different regions. To define the operator, let $\ket{a}$ be basis vectors for a state supported in $A$, and $\ket{b}$ be basis vectors for the state supported outside $A$. A basis for the entire system is then $\ket{a,b}$. For a given permutation $\pi$ on $R$ elements, we define the permutation operator $\pi$ on $R$ replicas of the Hilbert space using the action on basis states as
\begin{equation}
    \pi\left(\ket{a_1,b_1}\otimes ...\otimes\ket{a_i,b_i}\otimes...\otimes\ket{a_R,b_R}\right)=\ket{a_{\pi^{-1}(1)},b_1}\otimes ...\otimes\ket{a_{\pi^{-1}(i)},b_i}\otimes...\otimes\ket{a_{\pi^{-1}(R)},b_R}.
\end{equation}
That is, the permutation operator moves the vector, originally at $\pi^{-1}(i)$ to the replica of index $i$. Notice that this definition is independent of the choice of basis vectors, as for operator $U$ we have
\begin{equation}
    \pi\qty(\ket{a_1,b_1}\otimes ...\otimes\ket{Ua_{\pi^{-1}(i)},b_{\pi^{-1}(i)}}\otimes...\otimes\ket{a_R,b_R})=\ket{a_{\pi^{-1}(1)},b_1}\otimes ...\otimes\ket{Ua_{\pi^{-1}(i)},b_i}\otimes...\otimes\ket{a_{\pi^{-1}(R)},b_R}.
\end{equation}
so
\begin{equation}
    \pi\qty(\ket{Ua_1,b_1}\otimes ...\otimes\ket{Ua_i,b_i}\otimes...\otimes\ket{Ua_R,b_R})=\ket{Ua_{\pi^{-1}(1)},b_1}\otimes ...\otimes\ket{Ua_{\pi^{-1}(i)},b_i}\otimes...\otimes\ket{Ua_{\pi^{-1}(R)},b_R}.
\end{equation}
and we see that the permutation operator commutes with a basis change inside $A$ (and similarly inside $B$).
\section{Calculation of $\fabc$ for general string-net models}
\label{app: string-net}
\begin{figure}
  \begin{center}
  \begin{tikzpicture}

      \node at (0,0) {
  \begin{tikzpicture}
  \node[font={\normalsize}] at (-4.0,2.5) {(a)};
      \tikzset{hex/.style={semithick,gray,opacity=.2}}
      \def\r{.25};
      \newcommand{\hex}[2]{
      \draw[hex] ({#1+\r*cos(30)},{#2+\r*sin(30)}) \foreach \q in {90,150,...,390} {-- ({#1+cos(\q)*\r},{#2+sin(\q)*\r})};
      }
      \newcommand{\hexb}[2]{
      \draw[lightblue,very thick] ({#1+\r*cos(30)},{#2+\r*sin(30)}) \foreach \q in {90,150,...,390} {-- ({#1+cos(\q)*\r},{#2+sin(\q)*\r})};
      }
      \def\nx{2}
      \def\ny{2}
      \def\ss{1.73205080757}
      \foreach \x in {-7.5,-6.5,...,7.5}{
        \foreach \y in {-3.5,-2.5,...,3.5}{          \hex{sqrt(3)*\x*\r}{2*1.5*\y*\r}
            } }
      \foreach \x in {-8.0,-7.0,...,8.0}{
        \foreach \y in {-3.0,-2.0,...,3.0}{
            \hex{sqrt(3)*\x*\r}{2*1.5*\y*\r}
            }}
        \foreach \q in {0,120,240}{
        \begin{scope}[rotate=\q]
        \draw[semithick,opacity=.9,darkred] (-\ss*\r*3,9.0*\r) -- ++ (\ss*\r*5,0) --++ (-60:\ss*\r)--++(0:\ss*\r) -- ++ (-60:\ss*\r*5);
        \draw[semithick,opacity=.9,darkred] (0,0) -- (60:\ss*\r*5);
        \end{scope}}
        \foreach \t in {0,120,240}{
        \begin{scope}[rotate=\t]
            \foreach \x in {1,...,4}
                \hexb{-\x*\ss*\r}{0};
            \foreach \x in {1,...,6}
                \hexb{\ss*\r*(-6+\x/2)}{3*\r*\x/2};
            \foreach \x in {1,...,5}
                \hexb{\ss*\r*(\x-3)}{9*\r};
        \end{scope}}
        \draw[very thick,lightblue] (-\ss*\r,\r) to[out=120,in=-120] node[left] {$j$} (120:3*\ss*\r) to[out=0,in=120] node[right] {$k$} (0,2*\r);
        \draw[very thick,lightblue] (30:2*\r) to[out=0,in=120] node[above] {$k$} (0:3*\ss*\r) to[out=-120,in=0] node[below] {$m$} (-30:2*\r);
        \draw[very thick,lightblue] (-90:2*\r) to[out=-120,in=0] node[right] {$m$} (-120:3*\ss*\r) to[out=120,in=-120] node[left] {$j$} (-150:2*\r);
        \draw[very thick,lightblue] (-120:3*\ss*\r) to[out=-60,in=140]node[below] {$i$} (2*\ss*\r,-8*\r);
        \draw[very thick,lightblue] (120:3*\ss*\r) to[out=60,in=-140]node[below] {$n$} (2*\ss*\r,8*\r);
        \draw[very thick,lightblue] (120:3*\ss*\r) to[out=60,in=-140]node[below] {$n$} (2*\ss*\r,8*\r);
        \draw[very thick,lightblue] (0:3*\ss*\r) to[out=60,in=-120]node[right] {$l$} (3*\ss*\r,7*\r);
        \draw[very thick,lightblue] (2*\ss*\r,10*\r) to[out=60,in=90]node[right] {$n$} (7.5*\r*\ss,0);
        \draw[very thick,lightblue] (2*\ss*\r,-10*\r) to[out=-60,in=-90]node[right] {$i$} (7.5*\r*\ss,0);
        \draw[very thick,lightblue] (4*\ss*\r,8*\r) to[out=30,in=160]node[right] {$i$} (7.5*\r*\ss,0);
  \end{tikzpicture}};
  \node at (8,0) {
    \begin{tikzpicture}[every node/.style={black,font={\scriptsize}}]
    \node[font={\normalsize}] at (0,1.5) {(b)};
      \tikzset{lat/.style={very thick,lightblue}}
        \draw[thick,darkred,opacity=.9] (-.5,0) -- (4.5,0);
        \draw[lat] (0,-.5) --node[left,pos=.7] {$q_n$}++(0,1);
        \draw[lat] (0,-.5) --++ (1,0) -- node[below] {$r_{n-2}$}++ (.5,0);
        \draw[lat] (0,.5) --++ (1,0) -- node[above] {$r_{n-2}$}++ (.5,0);
        \draw[lat] (4,-.5) -- node[below] {$r_1$} ++ (-1,0) -- node[below] {$r_2$} ++ (-.5,0);
        \draw[lat] (4,.5) -- node[above] {$r_1$} ++ (-1,0) -- node[above] {$r_2$} ++ (-.5,0);
        \draw[lat] (4,-1.5)  --node[right] {$q_0$} ++(0,1) -- node[right,pos=.7] {$q_1$}++ (0,1) --node[right] {$q_0$}++ (0,1) ;
        \draw[lat] (1,-.5) -- node[right,pos=.7] {$q_{n-1}$} ++ (0,1);
        \draw[lat] (3,-.5) -- node[right,pos=.7] {$q_2$} ++ (0,1);
        \draw[lightblue,dotted,very thick] (1.5,-.5) --++ (1,0);
        \draw[lightblue,dotted,very thick] (1.5,.5) --++ (1,0);
        \node[font={\Large},darkred] at (0,-1.2) {$\pi$};
    \end{tikzpicture}};
    \end{tikzpicture}
  \end{center}
  \caption{(a) The simplest string-net model for the calculation of $\Phi$. Each pair of regions is connected by an extensive (area-law) number of strings, but the total ``charge" passing between the regions is well-defined (these are the indices $i,j,k,l,m,n$). Any string-net model can be converted to such configuration using unitaries that act within the regions and therefore preserve $\Phi$. (b) the summation over the configurations of a single ladder, given the total charge $q_0$ passing through it, allows us to reduce the diagram to a tetrahedron which describes only the total anyonic charge passing between the regions.}\label{fig:bubbles}
\end{figure}
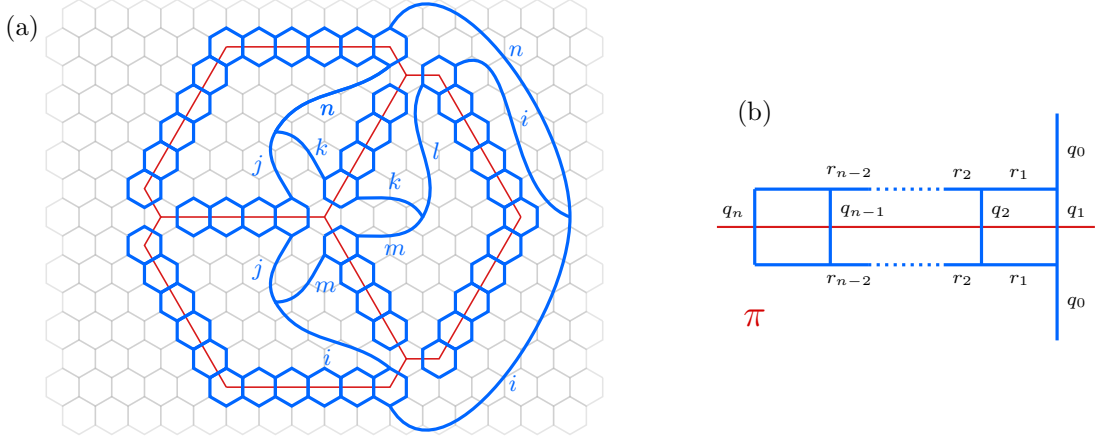

Here we generalize the calculation of $\fabc$ from abelian string-net models to more general (possibly non-abelian) models. We begin with a string-net model defined on some lattice, partitioning the lattice as in Fig. \ref{fig:pizza}. Since $\fabc$ is invariant under the action of any unitary which acts inside the regions $A,B,C$, we can disentangle the lattice sites within the region, to obtain a string-net state on a reduced graph, as depicted in Fig. \ref{fig:bubbles}a. The reduced graph still contains an extensive number of bonds passing between each pair of regions, but the information on the inside of each region is contained only in the total charge passing between neighboring regions, labeled by the edges $i,j,k,l,m,n$ in Fig. \ref{fig:bubbles}a. The form of the resulting graph is akin to Fig. \ref{fig: reduced-tetrahedron}, but with additional ``ladder" diagrams on the interface between any two regions. These ladder diagrams contain the area-law contribution to $\fabc$.

Our first task is to reduce the ladder diagram between any two regions, such that the summation will be over the tetrahedral graph. Let us calculate the contribution, coming from the ladder alone, to the expectation value of applying replica permutations. Consider the ladder diagram with $n$ edges between two regions $I,J$ with a cyclic permutation $\pi$ acting on $R$ replicas on region $I$, as in Fig. \ref{fig:bubbles}b. We assume that a total anyonic charge $q_0$ passes between the regions that are separated by the ladder. Summing over all possible configurations of the ladder degrees of freedom $\qty{q_i,r_i}$, we have 
\begin{equation}
    \Theta_{q_0}=\mel{\phi_{q_0}\ncp{R}}{\pi}{\phi_{q_0}\ncp{R}} =\frac{1}{D^{2R(n-1)}}\sum_{\qty{q,r}}N_{q_0 r_1}^{q_1}N_{r_1 q_2}^{r_2}\cdots N_{r_{n-2}q_{n-1}}^{q_n}d^{-R}_{q_0}\prod_{i\ge1} d_{q_i}^R,
\end{equation}
where we set $D^2=\sum_i d_i^2$ (notice that this is the quantum dimension of the fusion category used to construct the string net model, the total quantum dimension of the resulting topological order will be $D_{\rm doubled}=D^2$). In the limit of $n\to \infty$ we can set $N_{ab}^c\approx \frac{d_a d_b d_c}{D^2}$ \cite{flammia_topological_2009}. There are $n-1$ $N$ matrices, and we sum over $n-2$ $r_i$ indices, so we obtain
\begin{equation}
\begin{aligned}
  \Theta_{q_0} &\approx d_{q_0}^{1-R}\frac{1}{D^{2Rn-2(R-1)}}\sum_{\qty{q}} \prod_{i\ge 1}d_{q_i}^{R+1}\\
  &=S_R^n \frac{D^{2(R-1)}}{d_{q_0}^{R-1}}
\end{aligned}
\end{equation}
where we set
\begin{equation}
    S_R=\frac{\sum_i d_i^{R+1}}{D^{2R}}.
\end{equation}
If $\pi$ is replaced by a permutation that is not cyclic, a similar calculation will give 
\begin{equation}
  \Theta_{q_0}^\pi = \qty(\prod_c S_{R_c})^n\qty(\frac{D^2}{d_{q_0}})^{\abs{\pi}-R}
  \label{eq: Lambda}
\end{equation}
where $\abs{\pi}$ is the number of disjoint cycles of $\pi$ (or the minimal number of cyclic permutation that comprise $\pi$), and $R_c$ are the lengths of the disjoint cycles. In particular, we have $\abs{\pi_A}=2,\abs{\pi_B}=\abs{\pi_C}=r$. Intuitively, the $\Theta$ factors stem from the fact that for larger $d_{q_0}$ there are more ways for the charge $q_0$ to go between the two regions. Notice that in the case of non-abelian string-net models, it is necessary to start with a large number of edges and reduce the ladder, since, similarly to the case of the Rényi entropy \cite{flammia_topological_2009}, the value of $\Phi(r)$ obtains corrections that decay exponentially with the size of the regions.

The above calculation allows us to reduce the calculation of $\fabc$ to the summation over tetrahedron diagrams that describe the total anyonic charge going between the regions, as depicted in Fig. \ref{fig: reduced-tetrahedron}. A single tetrahedron diagram gives \cite{levin_string-net_2005}
\begin{equation}
  G^{ijm}_{kln}\equiv\vcenter{\hbox{
    \begin{tikzpicture}[every path/.append style={thick}, every node/.append style={font={\footnotesize}}]
    \def\r{1.1}
    \coordinate (a) at (0,0) {};
    \coordinate (b) at (90:\r) {};
    \coordinate (c) at (210:\r+.4) {};
    \coordinate (d) at (330:\r+.4) {};
    \draw[midarr] (b) -- node[right] {$i$} (a);
    \draw[midarr] (c) -- node[above] {$j$} (a);
    \draw[midarr] (d) -- node[above] {$m$} (a);
    \draw[midarr] (c) -- node[below] {$k$} (d);
    \draw[midarr] (b) -- node[right] {$l$} (d);
    \draw[midarr] (b) -- node[left] {$n$} (c);
    \end{tikzpicture}}
    }=F^{ijm}_{kln}\sqrt{d_i d_j d_k d_l},
  \label{eq: tetrahedron}
\end{equation}
where $F^{ijn}_{klm}$ are the $F$-symbols of the theory which defines the string-net model. The tetrahedral diagram gives the (not gauge invariant) information about the twists of the topological theory. We will soon see that the sum defining $\fabc$ allows the extraction of gauge-invariant information from the tetrahedral diagrams.

To carry out the sum, we notice that $\ket{\y}$ is a sum of all possible string configurations, with amplitudes obtained from the appropriate anyon diagrams (after reduction to the tetrahedral graph). The action of a permutation requires that each element in the superposition $\ket\rpsi$ matches an element in the superposition $\bra\rpsi$. In particular, it requires that if a configuration of strings $a_{s,t}$ passes between regions $i,j$, we must sum only over configurations for which $a_{\pi_i(s,t)}=a_{\pi_j(s,t)}$. That is, after cutting and permuting the regions, we require the anyon lines to match. This constrains the anyon lines that should be summed over. Explicitly, we have 
\begin{equation}
  \begin{aligned}
    i_{s,t}&=i_{s,t'}\equiv I_s \\
    n_{1,t}&=n_{2,t}\equiv N_t \\ 
    l_{1,t}&=l_{2,t-1}\equiv L_t \\ 
    k_{s,t}&=k_{s,t'}\equiv K_s \\ 
    j_{1,t}&=j_{2,t-1}\equiv J_t \\ 
    m_{1,t}&=m_{2,t-2}\equiv M_t
  \end{aligned}
  \label{eq: anyon-lines-summation}
\end{equation}
The sum comprising $\fabc$ then restricts to configurations matching the requirements \eqref{eq: anyon-lines-summation}. We obtain 
\begin{equation}
  \begin{aligned}
    \fabc =\Sigma_{\qty{L}} D^{12r}\sum_{\qty{IJKLMN}}   \Bigg(\prod_{t=1}^r& G^{I_1 J_t M_t}_{K_1L_nN_n}G^{I_2J_{t+1}M_{t+2}}_{K_1L_{t+1}N_t} \qty(G^{I_1J_{t+1}M_{t-1}}_{K_2L_tN_t})^*\qty(G^{I_2J_tM_{t-1}}_{K_1 L_{t+1}N_t})^* \Bigg)\\
     &\times \qty(\prod_{s=1,2} d_{I_s}d_{K_s})^{1-r}\qty(\prod_{t=1,\dots,r}d_{J_t}d_{L_t}d_{M_t}d_{N_t})
\end{aligned}
  \label{eq: phi123-non-abelian}
\end{equation}
Here the sum is taken over all configurations of the indices defined in \eqref{eq: anyon-lines-summation}. The two unconjugated $G$ factors are obtained from the first and second layers of $\ket\rpsi$, while the indices of the two conjugate $G$'s are obtained by acting with the replica permutation operators on the indices of the first two $G$ factors. The quantum dimension factors in the last lines are obtained from \eqref{eq: Lambda}. The term $\Sigma$ captures the area-law contributions, and is given by
\begin{equation}
    \Sigma_{\qty{L}}=S_{2}^{r\qty(L_{B\Lambda}+L_{C\Lambda}+L_{AB}+L_{BC})}S_r^{2\qty(L_{A\Lambda}+L_{BC})}
\end{equation}
where $L_{ij}$ is the length of the boundary between regions $i,j$. We can gain additional intuition about the area-law term by considering abelian models. In this case we obtain
\begin{equation}
    \Sigma_{\qty{L}}=D^{-2r\qty(L_{B\Lambda}+L_{C\Lambda}+L_{AB}+L_{BC})-4(r-1)\qty(L_{BC}+L_{A\Lambda})}.
\end{equation}
We notice that as the size of the regions increase, the boundaries that contribute the most to the area law are $A\Lambda$ and $BC$, and the area-law contribution decreases like $e^{-(r-1)L/\xi}$, obtaining the scaling in \eqref{eq: phi-def}. 

We verified numerically that \eqref{eq: phi123-non-abelian} gives the expected result for \eqref{eq: phi-mag}. The sum is generally difficult to evaluate as it contains exponentially many terms in $r$ (at least naively). We evaluated the sum for small values of $r$, and indeed found that the correct values are obtained when the string-net describes doubled Fibonacci ($r=2,...,6$), doubled Ising ($r=2,3$), and $D(S^3)$ ($r=2,3$) theories. In all cases, the numerical results reproduce \eqref{eq: phi-def} and \eqref{eq: phi-mag} exactly (to machine precision). For abelian models, the sum \eqref{eq: phi123-non-abelian} reduces to \eqref{eq: phi-abelian} when $n=1$, and gives additional area-law contributions when $n>1$, verifying the validity of the reduction to the reduced tetrahedron presented in the main text.

\section{Description of twisted $\Z_N$ gauge theory}
\label{app: Zn twisted gauge}
Here we describe the structure of twisted $\Z_N$ gauge theory, focusing on the topological spins. Twisted $\Z_n$ gauge theories can be described as abelian Chern-Simons gauge theories, with the $K$ matrix 
\begin{equation}
  K=\begin{pmatrix} 0 & N \\ N & -2p \end{pmatrix},
  \label{eq: twisted-zn-k-matrix}
\end{equation}
where $p=0,...,N-1$. For $p=0$ this gives the $\Z_N$ toric code. The anyons in the theory can be described as tuples $(s,m)$ of ``magnetic" and ``electric" charges, with $s,m=0,...,N-1$ and the fusion rules $(s_1,m_1)\times (s_2,m_2)=(s_1+s_2,m_1+m_2)$ (with addition mod $N$). The spins of the anyons are given by 
\begin{equation}
  \theta_{(s,m)} = \exp[2\pi i \qty(\frac{p s^2}{N^2}+\frac{ms}{N})],
  \label{eq: zn_t_mat}
\end{equation}
and the mutual statistics (or the $S$-matrix) is given by 
\begin{equation}
    \theta_{(s,m),(t,n)} = \exp[2\pi i \qty(\frac{2pst}{N^2}+\frac{sn+tm}{N})].
    \label{eq: zn_s_mat}
\end{equation}
As a sidenote, we mention that for systems with charge conservation the theory also has a charge vector $t$, such that $t^TK^{-1}t$ describes the Hall conductivity. In that case, different values of $p$ necessarily correspond to different quantum phases. In contrast, in the absence of charge conservation, theories with $p'=pk^2$ for $k \in \mathbb{Z}$ are the same [up to permutation of the anyon labels $(s,t)$], as they have the same $S$ and $T$ matrices up to relabeling of the anyons [this can be verified using \eqref{eq: zn_t_mat}, \eqref{eq: zn_s_mat}].

Next, we want to calculate the sums over the topological spins, given by 
\begin{equation}
  \begin{aligned}
    N\Phi(r)=\sum_a d^2_a \theta_a^r&=\sum_{s,m}\exp{2\pi ir \qty(\frac{p s^2}{N^2}+\frac{ms}{N})} \\ 
    &= N\sum_{s|\frac{N}{g}}\exp(2\pi ir \frac{p s^2}{N^2}) \\ 
    &= N\sum_{q=0}^g \exp(2\pi ir \frac{p q^2}{g^2}) ,
  \end{aligned}
  \label{eq: Z_N spins}
\end{equation}
where $g=\gcd(N,r)$ and $s|\frac{N}{g}$ denotes ``$s$ divides $\frac{N}{g}$". In the case where $r|N$ this gives $g=r$ so 
\begin{equation}
  \Phi(r)= \sum_{q=0}^r \exp(2\pi i \frac{p q^2}{r}).
\end{equation}

\section{Derivation of \eqref{eq: phi-abelian} for general $r$}
\label{app: abelian general r}
Here we derive $\fabc$ for a twisted $\Z_N$ theory and general $r$, removing the assumptions using in the main text, that $N$ is prime and $r=N$. In the more general case, \eqref{eq: phi-abelian} becomes
\begin{equation}
  \begin{aligned}
    \fabc&=\frac{1}{N^{6r}}\sum_{A,B,C,\delta} \prod_{t=1}^rF(A,B+\td,C)F(A+\td,B+(t+1)\td,C+\td)F(A,B+(t+1)\td,C+\td)^* F(A+\td,B+t\td,C)^* \\ 
            &=\frac{1}{N^{6r}}\sum_{A,B,C,\delta} \exp\qty{\frac{2\pi i p\td}{N} \sum_t\qty[\qty{B+(t+1)\td,C+\td}-\qty{B+t\td,C}]} \\ 
            &=\frac{1}{N^{6r-1}}\sum_{B,C,\delta} \exp\qty{\frac{2\pi i p\delta}{g} \sum_t\qty[\qty{B+t\frac{\delta N}{g},C+\frac{\delta N}{g}}-\qty{B+t\frac{\delta N}{g},C}]},
\end{aligned}
  \label{eq: phi-abelian-general}
\end{equation}
where we recall that $\tilde{\delta}=\frac{N}{g}\delta$. On the other hand, we have 
\begin{equation}
    \sum_{t=1}^r \qty{b+t\frac{\delta N}{g},c} =\sum_{t=1}^r \qty{b+t\delta^*,c}  =\frac{r\delta^*}{N}\qty(\qty(c:\delta^*)+ 1_{c^*+b^*>\td}),
\end{equation}
where $\delta^*=\gcd(\tilde{\delta},N)$, $c^*$ is the reminder of $c$ divided by $\delta^*$, and $c:\delta^*=(c-c^*)/\td$. Eq. \eqref{eq: phi-abelian} then gives 
\begin{equation}
  \begin{aligned}
    \fabc &= \frac{1}{N^{6r-1}}\sum_{B,C,\delta} \exp\qty(\frac{2\pi i p \delta}{g}\times\frac{r\delta^*}{N}\times\frac{\delta N}{g\delta^*}) \\ 
    &=\frac{1}{N^{6r-3}}\sum_{\delta=0}^g \exp\qty(\frac{2\pi i p r \delta^2}{g^2}),
  \end{aligned}
\end{equation}
which is the expected result [see Eq. \eqref{eq: Z_N spins}].

\section{Impossibility of extracting invariants of topological order from a single replica}
Here we show that, in the absence of symmetries, there is no operator $O$, acting on a single replica, whose expectation value $\mel{\psi}{O}{\psi}$ can distinguish between two different topological phases (in the sense that $\ev{O}$ gives a set of values on wavefunctions in phase 1, and a different set of values for wavefunctions in phase 2). The argument should be seen as a toy version of the one presented in App. \ref{app: operator optimality proof} for distinguishing between different twisted $\mathbb{Z}_n$ gauge theories.

We consider the set of random unitaries
\begin{equation}
    U=\prod_i U_i
\end{equation}
where $U_i$ are single-qubit Haar-random unitaries. Clearly, $U$ are finite-depth unitaries, so all of $U\ket{\psi}$ will be of the same topological phase. We decompose $O$ in the Pauli basis as $O=\sum_p o_p P$, where $P$ is a Pauli string. We can assume that $O$ has no unity component, namely $o_{I}=0$. We obtain
\begin{equation}
\begin{aligned}
    \mathbb{E}_U \mel{\psi}{U^\dagger OU}{\psi} &=\sum_p o_p \mathbb{E}_U \mel{\psi}{U^\dagger PU}{\psi} \\
    &=\sum_p o_p \mathbb{E}_U \mel{\psi}{U^\dagger U^\dagger_p PU_pU}{\psi} \\
    &=-\sum_p o_p \mathbb{E}_U \mel{\psi}{U^\dagger PU}{\psi}=0 ,
\end{aligned}
\end{equation}
where, for each $P$, we chose a single-qubit unitary $U_p$ such that $\qty{U_p,P}=0$. Since this expectation value is independent of $\psi$, $\ev{O}$ cannot distinguish between different topological phases.

Notice that the unitaries $U$ break all symmetries of $\psi$, including translation symmetry. If translation symmetry is preserved, there exist methods to extract universal data from a single replica, for example \cite{tu2013momentum, kobayashi_extracting_2024}.
\section{Proof of Conjecture 1 for a tetrahedral lattice}
\label{app: operator optimality proof}
\begin{figure}
    \centering
    \begin{tikzpicture}[every path/.append style={thick}, every node/.append style={font={\scriptsize}}]
    \def\r{1.5}
    \coordinate (a) at (0,0) {};
    \coordinate (b) at (90:\r) {};
    \coordinate (c) at (210:\r+.4) {};
    \coordinate (d) at (330:\r+.4) {};
    \draw[midarr] (b) node[right] {$v_\delta$} -- node[right] {$a$} (a);
    \draw[midarr] (c) node[left] {$v_\gamma$} -- node[above] {$b$} (a)  node[right] {$v_\alpha$};
    \draw[midarr] (a) -- node[above] {$a+b$} (d);
    \draw[midarr] (c) -- node[below] {$c$} (d) node[right] {$v_\beta$};
    \draw[midarr] (d) -- node[right] {$a+b+c$} (b);
    \draw[midarr] (b) -- node[left] {$b+c$} (c);
    \end{tikzpicture}
    \caption{The tetrahedral graph.}
    \label{fig: app-tetra}
\end{figure}
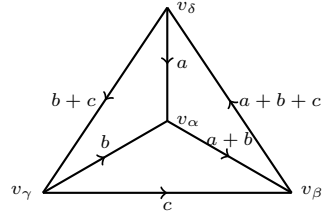

\newcommand{\du}{\mathcal{D}U}
\newcommand{\ps}[1]{\ket{\psi_{#1}}}
\newcommand{\ull}{U_{\qty{\ell_i,\ell_i'}}}
Here we prove \eqref{eq: fooling-o} for the unitaries defined in \eqref{eq: u-vertex-decomposition} and $k<2N$, assuming the simple lattice geometry in Fig. \ref{fig: app-tetra}. The form of the wavefunctions is given by \eqref{eq: psi-string-net}. The two wavefunctions $\ket{\psi_i}$ then differ by having the different phase factors for the same string configurations. For simplicity, we can assume that $\ket{\psi_1}$ corresponds to the untwisted $\Z_N$ toric code, so $F_1=1$ (where $F_i$ are the $F(a,b,c)$ symbols presented in 
\eqref{eq: psi-string-net} corresponding to the wavefunctions $\ket{\psi_{i=1,2}}$), and we'll further denote $F_2\equiv F$.

Since $\ps{1,2}$ differ from each other only by the phase factors $F$, it makes sense to choose the random circuits $U$ to include phase gates only. We recall that $U$ are defined in the main text by
\begin{equation}
\begin{aligned}
  U&=\prod_v U_v, \\
    U_v\ket{
      \vcenter{\hbox{
      \begin{tikzpicture}[every path/.style={thick}, every node/.style={font={\scriptsize}},scale=.7]
        \draw[thick,midarr] (-30:1) --node[above] {$a$} (0,0);
        \draw[thick,midarr] (-30+240:1) --node[above] {$b$} (0,0);
        \draw[thick,midarr] (0,0) --node[right] {$a+b$} (0,1);
  \end{tikzpicture}}}
      }&=e^{i \phi_{ab}}\ket{
      \vcenter{\hbox{
      \begin{tikzpicture}[every path/.style={thick}, every node/.style={font={\scriptsize}},scale=.7]
        \draw[thick,midarr] (-30:1) --node[above] {$a$} (0,0);
        \draw[thick,midarr] (-30+240:1) --node[above] {$b$} (0,0);
        \draw[thick,midarr] (0,0) --node[right] {$a+b$} (0,1);
  \end{tikzpicture}}}
};& \phi_{ab}&\in\qty[0,2\pi].
  \end{aligned}
  \label{eq: u-def-app}
\end{equation}
with $\phi_{ab}$ on each vertex chosen uniformly and independently at random.
Importantly, the random measure on $\qty{U}$ is a Haar measure, so, for any $U_0$ of the form \eqref{eq: u-def-app}, we have 
\begin{equation}
  \int \du\qty(U^\ok)^\dagger \mathcal{O} U^\ok =
  (U_0^\ok)^\dagger\qty(\int \du\qty(U^\ok)^\dagger \mathcal{O} U^\ok )U_0^\ok.
  \label{eq: haar}
\end{equation}
Using \eqref{eq: psi-string-net} and \eqref{eq: haar} we can expand \eqref{eq: fooling-o} as (we use $\ell_i$ as a shorthand for $(a_i,b_i,c_i)$)
\begin{equation}
  \begin{aligned}
  &\int\du \mel{\psi_1^\ok}{\qty(U^\ok)^\dagger \mathcal{O}U^\ok}{\psi^\ok_1}\\
  =&\int\du \sum_{\qty{\ell_i,\ell_i'}}\mel{\ell_1,..., \ell_k}{\qty(U^\ok)^\dagger \mathcal{O}U^\ok}{\ell_1', ..., \ell_k'}\\
  =&\int\du \sum_{\qty{\ell_i,\ell_i'}}\mel{\ell_1,...,\ell_k}{\qty(U^\ok_{\qty{\ell_i,\ell_i'}})^\dagger \qty(U^\ok)^\dagger \mathcal{O}U^\ok U^\ok_{\qty{\ell_i,\ell_i'}}}{\ell_1',...,\ell_k'},
  \end{aligned}
  \label{eq:fooling-o-expanded}
\end{equation}
where $U^\ok_{\qty{\ell_i,\ell_i'}}$ is any choice of unitary that depends on $\qty{\ell_i,\ell_i'}$. A sufficient condition for \eqref{eq: fooling-o} is, therefore, that for any $\ell_1,...,\ell_k,\ell_1',...,\ell_n'$ we can choose $\ull$ such that 
\begin{equation}
  \begin{aligned}
  &\ull^\ok\otimes(\ull^*)^\ok\ket{\ell_1,...,\ell_k,\ell_1',...,\ell_k'} \\
    =&F(\ell_1)\cdots F(\ell_k)
    F(\ell'_1)^* \cdots F(\ell_k')^*\ket{\ell_1,...,\ell_k,\ell_1',...,\ell_k'}.
  \end{aligned}
  \label{eq: phase-u-choice}
\end{equation}
Here the two unitaries in the first line act on the first and second sets of $\ell_i$, and we replaced the bra in \eqref{eq:fooling-o-expanded} with an extended ket that includes $2k$ configurations. We now ask what is the minimal $k$ such that $\ull$ cannot be chosen to satisfy \eqref{eq: phase-u-choice}. A necessary and sufficient condition is that there exists a configuration of strings $\qty{\ell_i,\ell_i'}$ for which 
\begin{equation}
  F(\ell_1)\cdots F(\ell_k)F^*(\ell_1')\cdots F^*(\ell_k')\neq 1,
  \label{eq: f-prod-condition}
\end{equation}
and, simultaneously, for $U$ of the form \eqref{eq: u-vertex-decomposition},\eqref{eq: uv-req} we have
\begin{equation}
  \begin{aligned}
  &U^\ok\otimes(U^*)^\ok\ket{\ell_1,...,\ell_k,\ell_1',...,\ell_k'} = \ket{\ell_1,...,\ell_k,\ell_1',...,\ell_k'}.
  \end{aligned}
  \label{eq: minimal-strings}
\end{equation}
That is, no $U$ can affect the phase of the string configuration [the requirement \eqref{eq: minimal-strings} is necessary since if the phase on the RHS is not 1 we can always raise $U$ to some power to obtain the phase in \eqref{eq: f-prod-condition}]. Using the definition \eqref{eq: u-def-app}, this requirement can be rephrased as follows: for each vertex $v$ in the configuration $\ell_i$, we need to ``cancel" the vertex contribution by another configuration $\ell_j'$, that has the same strings around $v$. Also, for a configuration with minimal $k$ we can assume that $\ell_i\neq\ell_j'$ for any $i,j$, otherwise we could remove both to obtain a solution with smaller $k$.

Up to this point, the argument was generic for any edge lattice. To finalize the argument for the tetrahedral lattice presented in Fig. \ref{fig: app-tetra}, we need to show that any configuration satisfying \eqref{eq: f-prod-condition} and \eqref{eq: minimal-strings} has $k\ge2N$. Assuming without loss of generality that $F(\ell_1)\neq1$, in particular $a_1,b_1,c_1\neq0$ [this can be seen from \eqref{eq:Z_N-f-sym}]. The condition \eqref{eq: minimal-strings} requires that there is another string configuration, say $\ell_1'$, for which $v_\alpha$ is the same vertex as in $\ell_1$ (so $a_1,b_1=a_1',b_1'$ are the same) but $\ell_1\neq\ell_1'$, so $c_1\neq c_1'$. In that case, we must have a different string configuration, say $\ell_2$, with the same $v_\beta$ vertex as $\ell_1'$, so $c_2=c_1',a_2+b_2=a_1'+b_1'$ but $a_2\neq a_1$. Going on, we get a set of string configurations $\ell_i,\ell_i', i=1,...,\tau$ satisfying
\begin{equation}
  \begin{aligned}
    a_i+b_i &= a_i'+b_i' =a_{i+1}+b _{i+1} \\
    c_i'&\neq c_i \\
    a_i&\neq a_{i+1} \\
    c_{i+1}&=c_i' 
  \end{aligned}
  \label{eq: sun-diagram-condition}
\end{equation}
(where $i+1$ implies addition mod $\tau$). The simplest guess satisfying \eqref{eq: sun-diagram-condition} (which, as we show, does not work) has $\tau=2$ and is given generally by
\begin{equation}
  \begin{aligned}
    (a_1,b_1,c_1) &= (a,b,c), \\
    (a_1',b_1',c_1') &= (a,b,c+q), \\
    (a_2,b_2,c_2) &= (a+p,b-p,c+q), \\
    (a_2',b_2',c_2') &= (a+p,b-p,c),
  \end{aligned}
  \label{eq: configs 1}
\end{equation}
 with $p,q\neq0$. This guess ensures that \eqref{eq: minimal-strings} is satisfied when $U$ is nontrivial only on the vertices $v_\alpha,v_\beta$, but fails in general since these configurations give different values for $v_\gamma$ and $v_\delta$, which must also be canceled. That is, for each configuration with $(a,b,c)$ we need another configuration of the form $(a+s,b,c)$. Similarly, to cancel the contribution from $v_\delta$, for each configuration of the form $(a,b,c)$ we need another of the form $(a,b+s,c-s)$. We can satisfy all requirements by letting $b$ run over an additional index, such that
\begin{equation}
  \begin{aligned}
  \ell_{1,t}\equiv(a_{1,t},b_{1,t},c_{1,t}) &= (a,b+tp,c), \\
  \ell'_{1,t}\equiv(a_{1,t}',b_{1,t}',c_{1,t}') &= (a,b+tp,c+p), \\
  \ell_{2,t}\equiv(a_{2,t},b_{2,t},c_{2,t}) &= (a+p,b+(t-1)p,c+p), \\
  \ell'_{2,t}\equiv(a_{2,t}',b_{2,t}',c_{2,t}') &= (a+p,b+(t-1)p,c).
  \end{aligned}
  \label{eq: string-req-sol}
\end{equation}
These are copies of the configurations \eqref{eq: configs 1}, so the requirements on $v_1,v_2$ are satisfied automatically. The cancellation of $v_3$ occurs between $\ell_{1,t}$ and $\ell'_{2,t+1}$, and between $\ell_{2,t}$ and $\ell_{1,t+1}'$. The cancellation of $v_4$ occurs between $\ell_{1,t}$ and $\ell_{1,t-1}'$, and between $\ell_{2,t}$ and $\ell_{2,t+1}'$. Importantly, we must let the index $t=1,...,N$ as $a,b,c$ are defined mod $N$. This gives $k=2N$ string configurations, exactly as one given in the statement above \eqref{eq: fooling-o}. Note that there are other possible configurations of the tetrahedral diagram, with the same number of string configurations, obtained by permutations of the vertex labels. We also comment that, if $N$ is not prime, we can choose $q$ such that $m=\gcd(q,N)>1$, in that case, we only need $k=2N/m$ replicas. 

Finally, it is worthwhile to discuss the generalization of the above result to lattices beyond the tetrahedral lattice described here. While we do not have a rigorous argument, we believe the result generalizes. The heuristic argument is as follows: On a larger lattice, one still considers string configurations solving \eqref{eq: f-prod-condition} and \eqref{eq: minimal-strings}. In larger string configurations, we similarly need all vertices to cancel between $\qty{\ell_i}$ and $\qty{\ell_i'}$. This will be harder to satisfy in a non-trivial way, and importantly, nothing is gained by considering configurations of disconnected strings (that is, in which edges on which the strings are non-zero form disconnected graphs), as such loop configurations factorize to the disconnected components. The upshot is that, even on larger lattices, configurations that solve the two requirements will differ only on a subgraph of the lattice that is of the topology of the tetrahedral graph, such that the differences between the string configurations are of the form of \eqref{eq: string-req-sol}.
\end{document}